\documentclass{article}
\usepackage{theorem}
\usepackage{graphicx}
\usepackage{fullpage}
\usepackage{amssymb,amsmath}
\usepackage{theorem}
{\theorembodyfont{\rmfamily}
\newtheorem{definition}{Definition}[section]}
\newtheorem{theorem}{Theorem}[section]
\makeatletter
\newcommand{\card}{\mathop{\operator@font card}}
\makeatother
\begin{document}
\frenchspacing
\title{Fuzzy Bigraphs:\\ An Exercise in Fuzzy Communicating Agents}
\author{Apostolos Syropoulos\\
       Greek Molecular Computing Group\\
       Xanthi, Greece\\
       \texttt{asyropoulos@yahoo.com}}
\maketitle
\begin{abstract}
Bigraphs and their algebra is a model of concurrency. Fuzzy bigraphs are a generalization of
birgraphs intended to be a model of concurrency that incorporates vagueness. More specifically,
this model assumes that agents are similar, communication is not perfect, and, in general,
everything is or happens to some degree. 
\end{abstract}
\section{Introduction}
In a way, the work of the ACM Turing Award laureate Robin Milner\cite{milner1989,milner1999} defined the evolution 
of concurrency theory. Initially, his CCS calculus and later on his $\pi$-calculus were milestones in the
development of concurrency theory. His work was an algebraic approach to concurrency and communication. Later on, 
Milner~\cite{jensen2003,milner2009} proposed the use of bigraphs, which combines graph theory with category theory, 
for the description of interacting mobile agents. In fact, Milner was interested in a mathematical model of ubiquitous 
computing, which subsumes concurrency theory. Although the use of category theory is a very interesting development, 
still the bigraphical model assumes that agents, processes, etc., are {\em crisp}. This means that two processes are 
either identical or different. It also means that communication happens or does not happen. Obviously, there is nothing 
wrong with approach but I would dare to say that this is not natural and to some degree not realistic! In general, two 
processes or agents can be either identical or different, nevertheless, it is always possible that they are similar to 
some degree. For example, think of two instances of a word processor running on a computer system. Obviously, these 
processes are not identical (i.e., they do not consume exactly the same computational resources) but they are not entirely 
different. The idea that processes can be similar to some degree has been discussed in detail by this author~\cite{syropoulos2014}. 
This idea is part of a more general thesis according to which vagueness is a fundamental property of our world. This means that 
there are vague objects, vague dogs, and vague humans. In philosophy this view is called {\em onticism}~\cite{akiba2014}. 
Personally, I favor this idea, but I do not plan to discuss the pros and cons of it here. Despite of this, I have not 
explained why category theory matters.

Category theory is a very general formalism with many applications in informatics. Instead of giving a {\em informal}
description of  category theory, I will quote Tom Leinster's~\cite{leinster2014} picturesque description of category 
theory: 
\begin{quote}
Category theory takes a bird's eye view of mathematics. From high in the sky,
details become invisible, but we can spot patterns that were impossible to detect
from ground level. How is the lowest common multiple of two numbers
like the direct sum of two vector spaces? What do discrete topological spaces,
free groups, and fields of fractions have in common?
\end{quote}
Mathematical structures (e.g., Hilbert spaces and Scott domains) are also used to describe physical 
and computational processes so category theory may give a bird's eye view of physics and computation. 
However, I am more interested in computation, in general, and fuzzy computation, in particular.

The theory of fuzzy computation employs crisp models of computation or crisp conceptual
computing devices to define vague models of computation and vague conceptual computing devices.
These models are defined by {\em fuzzifying} the corresponding crisp models. This may seem like
an oxymoron since in computation we are interested in exact results and here I am talking about 
vague computing. To resolve this problem, suffices to say that vague computing devices employ
vagueness to deliver an exact result. For example, the Hintikka-Mutanen TAE-machines~\cite{Syropoulos2018}
compute results in the limit by continuously printing ``yes'' and/or ``no'' on one of their tapes and
in the limit they print their final answer to the problem they are supposed to solve. Using vagueness
would mean that valid answers would include ``maybe'', ``quite possibly''`, etc. These answers could
be used to deliver the final answer easier as the machine does not {\em oscillate} between ``yes'' and
``no'' but approaches one of the two ends. Of course this is not a fully worked out model of computation
but it gives an idea of how vagueness is used in computation.

If we want to have a fuzzy version of Milner's bigraphs, we need to give a definition of fuzzy bigraphs. 
This definition should extend the definition of crisp bigraphs. Roughly a bigraph consists of a forest 
(i.e., a graph without any graph cycles) and a hypergraph (i.e., a graph 
in which edges, which are called {\em hyperedges}, may connect more than two nodes). Thus it is necessary to 
define fuzzy graphs and fuzzy hypergraphs. Fortunately, fuzzy graphs and fuzzy hypergraphs have been introduced 
by Azriel Rosenfeld~\cite{rosenfeld1975} and by William L. Craine~\cite{craine1993}, respectively. Milner's theory 
uses {\em precategories} and {\em s-categories}, which are like categories and partial monoidal categories, 
respectively, however they differ in that arrow composition is not always defined. To the best of my 
knowledge there are two fuzzy versions of category theory. In particular, Alexander {\v{S}}ostak~\cite{sostak1999} 
and this author~\cite{syropoulos2014b} presented two different definitions of fuzzy categories. Here we 
are going to use the later definition. 

\paragraph{Plan of the paper} First I will briefly explain basic notions of bigraph theory. Then, I will introduce
all the fuzzy mathematical structures that are required in order to give a fuzzy version of bigraphs. Next, I will
introduce fuzzy bigraphs and type~2 fuzzy bigraphs and I will a sketch of categories that have as arrows these
structures.

\section{Bigraphs in a Nutshell}
I expect readers to be familiar with basic notions from graph theory. However, I think most readers will not be
familiar with the notion of a hypergraph, which is a generalization of the concept of a graph. The
definition that follows is from~\cite{berge1976}:
\begin{definition}
Suppose that 
\begin{displaymath}
V=\bigl\{v_1,v_2,\ldots,v_n\bigr\}
\end{displaymath}
is a finite set and 
\begin{displaymath}
\mathcal{E}=\bigl\{E_i\mathrel{\bigm|}i\in I\bigr\}
\end{displaymath}
is a family of subsets of $V$. The
family $\mathcal{E}$ is said to be a {\em hypergraph on $V$} if
\begin{enumerate}
\item $E_i\neq\emptyset $ for all $i\in I$; and
\item $\bigcup_{i\in I} E_i=V$.
\end{enumerate}
\end{definition}

The pair $\mathcal{H}=(V,\mathcal{E})$ is called a {\em hypergraph}. The number  $n=\card V$ is called the
{\em order} of the hypergraph. The elements $v_1,\ldots,v_n$ are called the {\em vertices} and the sets
$E_1,\ldots,E_m$ are called the {\em edges}. Thus the big difference between a graph and a hypergraph is that
the edges of a hypergraph can be determined by one or more vertices while the edges of a graph are determined
always by two vertices.

\subsection{Informal Description of Bigraphs}

A bare bigraph consists of a forest (i.e., a graph that consists of trees) and a hypergraph. Their common
set of vertices or {\em nodes} is the set $V\subset\mathcal{V}$, where $\mathcal{V}$ is the infinite 
set of all possible nodes. On the other hand, the edges of the hypergraph form the set
$\mathcal{E}$. The set of vertices and the set of edges of the bigraph are the sets $V$ and $\mathcal{E}$,
respectively. Let us add some structure to these components. First, the trees that make up the forest should be 
rooted but also they should have designated terminal vertices (or nodes) that that are called {\em sites}. 
Such  a forest will be called a {\em place graph}. The hypergraph should have edges with missing endpoints.
These edges should be used to compose one hypergraph with another one. Such a hypergraph will be called a
{\em link graph}. A concrete bigraph is a pair consisting of a place graph and a link graph. 

A bigraph represents a snapshot of a ubiquitous computing system. A system represented by a bigraph can 
reconfigure itself and it can interact with its environment (e.g., other systems).
A graphical representation of a bigraph is shown in figure~\ref{bigraph::}. The nodes of a bigraph are used to 
encode real or virtual agents and are represented as ovals or circles. An agent can be a computer, a pad,
a smartphone, etc. The nesting of nodes describes their spatial placement. Interactions between agents are
represented by {\em links}. Each node can have zero, one or more {\em ports} (the bullets on the bigraph).
These ports are entry points and function just like the ports of a computer server that provides various
Internet services like SMTP at port 25, HTTP at port 80, etc.
Nodes are characterized by a {\em control}. Nodes that have the same control, have the same number of ports.
Dashed rectangles denote {\em regions} and are called {\em roots}. The roots specify adjacent parts of a system.
Shaded squares are called {\em sites}. They encode holes in a system that can be replaced with agents. A bigraph
can have inner and outer names (e.g., $y$ is an outer name and $x_1$, $x_2$ are inner names). These names encode
links (or potential links) to other bigraphs. The elements that make up a bigraph (i.e., nodes and edges) can be 
assigned unique identifiers, which is called the {\em support} of a bigraph. When a bigraphical structure has a 
support, it is called {\em concrete}. 

\subsection{Mathematical Description of Bigraphs}
\begin{figure}
\begin{center}
\includegraphics[scale=0.75]{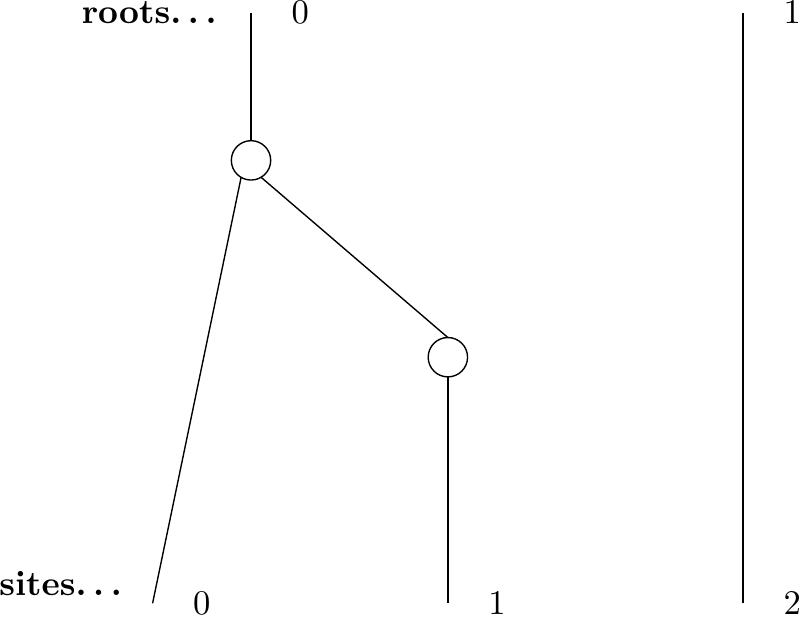}
\end{center}
\caption{The concrete place graph $H^{\mathsf{P}}:3\rightarrow 2$. Note that this is a forest consisting of two rooted trees.}\label{place::graph}
\end{figure}

\begin{figure}
\begin{center}
\includegraphics[scale=0.85]{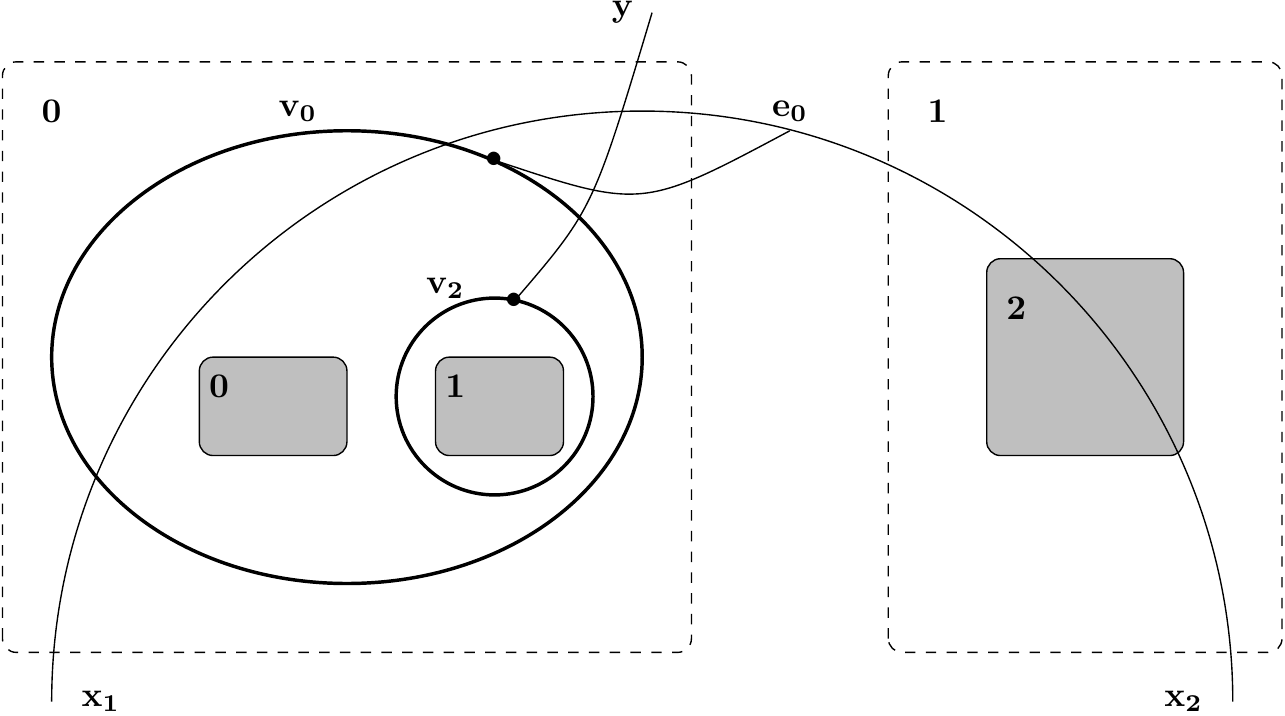}
\end{center}
\caption{The bigraph $H:\langle3,\{x_1,x_2\}\rangle\rightarrow\langle 2,\{y\}\rangle$.}\label{bigraph::}
\end{figure}

The notation  $f^{n}(x)$ is used to describe acyclic maps:
\begin{displaymath}
f^{n}(x)\; \text{denotes}\; \underbrace{f(f(f\cdots f(x)\cdots))}_{n\; \text{times}}.
\end{displaymath}
Also, $S\uplus T$ denotes the union of sets that are disjoint (i.e., $S\cap T=\emptyset$).
Before giving the formal definition of a concrete bigraph we need three auxiliary
definitions.
\begin{definition}
A {\em basic signature} is a pair $(\mathcal{K},\mathrm{ar})$, where $\mathcal{K}$ is a set of nodes
that are called {\em controls} and $\mathrm{ar}:\mathcal{K}\rightarrow\mathbb{N}$ is a function that
assigns an {\em arity} to each control.  
\end{definition}
For simplicity, when the arity is understood, the signature is written as $\mathcal{K}$. 
\begin{definition}
A concrete place graph 
\begin{displaymath}
F=(V_F,\mathrm{ctrl}_F,\mathrm{prnt}_F):m\rightarrow n
\end{displaymath}
is a triple  having an inner interface $m$ and an outer interface $n$, where $n$ and $m$ are 
{\em ordinal numbers}.\footnote{In set theory the natural number $0$ is defined to be the empty set, 
that is, $0\stackrel{\mathrm{def}}{=}\emptyset$. If $x$ is 
a natural number, then $x^{+}$ is its successor is defined as follows:
\begin{displaymath}
x^{+}\stackrel{\mathrm{def}}{=}x\cup\{x\}.
\end{displaymath}
Thus, numbers are identified with sets and so
\begin{align*}
1 &= 0^{+} = \{0\}=\{\emptyset\}\\
2 &= 1^{+} = \{0,1\}\\
3 &= 2^{+} = \{0,1,2,\}\\
\vdots &\phantom{=}\qquad \vdots\\
m &= (m-1)^{+} = \{0,1,2,\ldots,m-1\}
\end{align*}
The reader should consult any basic introduction to set theory  for more details (e.g., see~\cite{halmos1974}).}
These ordinals are used to enumerate the sites and the roots of the
place graph. $V_F\subset\mathcal{V}$ is the set of nodes, where $\mathcal{V}$ is an infinite set of node-identifiers, 
$\mathrm{ctrl}_F:V_F\rightarrow\mathcal{K}$ is a {\em control map}, 
and $\mathrm{prnt}_F:m\uplus V_F\rightarrow V_F\uplus n$ is called {\em parent map}. This map is acyclic, 
that is, if $\mathrm{prnt}_F^{i}(v)=v$ for some $v\in V_F$, then $i=0$. 
\end{definition}
Figure~\ref{place::graph} shows a concrete place graph.
\begin{definition}
A {\em concrete link graph}
\begin{displaymath}
F=(V_F,E_F,\mathrm{ctrl}_F,\mathrm{link}_F):X\rightarrow Y
\end{displaymath}
is a quadruple with inner interfaces $X$ and outer interfaces $Y$ that are finite subsets of $\mathcal{X}$,
where $\mathcal{X}$ is an infinite set of names. $X$ and $Y$ are called the {\em inner} and {\em outer names} 
of the link graph, respectively. $V_F\subset\mathcal{V}$ and 
$E_F$ is a finite subset of the infinite set $\mathcal{E}$ of edges.
Also, $\mathrm{ctrl}_F:V_F\rightarrow\mathcal{K}$ is a {\em control map}, 
and $\mathrm{link}_F:X\uplus P_F\rightarrow E_F\uplus Y$ is a {\em link map}, where
\begin{displaymath}
P_F\stackrel{\mathrm{def}}{=}\bigl\{(v,i)\mathrel{\bigm|}i\in\mathrm{ar}(\mathrm{ctrl}_F(u))\bigr\}
\end{displaymath}
is the set of ports of $F$. The pair $(v,i)$ denotes the $i$th port of vertex $v$. The sets
$X\uplus P_F$ and $E_F\uplus Y$ are the {\em points} (i.e., ports or inner names) and the 
{\em links} of F, respectively.
\end{definition}  
\begin{figure}
\begin{center}
\includegraphics[scale=0.75]{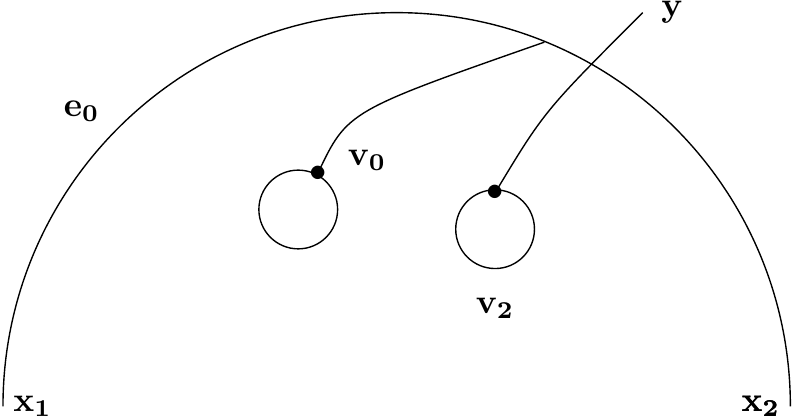}
\end{center}
\caption{The concrete link graph $H^{\mathsf{L}}:\{x_1,x_2\}\rightarrow\{y\}$. The bullets on 
vertices are ports.}\label{link::graph}
\end{figure}
Figure~\ref{link::graph} depicts a concrete link graph.
Now we can proceed with the definition of a concrete bigraph: 
\begin{definition}
An {\em interface} for bigraphs is a pair $I=\langle m,X\rangle$ of a place graph and a link
graph interface. The ordinal $m$  is the {\em width} of $I$. A {\em concrete bigraph}
\begin{displaymath}
F=(V_F,E_F,\mathrm{ctrl}_F,\mathrm{prnt}_F,\mathrm{link}_F):\langle k,X\rangle\rightarrow
                                                             \langle m,Y\rangle
\end{displaymath}
consists of a concrete place graph $F^{\mathsf{P}}=(V_F,\mathrm{ctrl}_F,\mathrm{prnt}_F):k\rightarrow m$
and a concrete link graph $F^{\mathsf{L}}=(V_F,E_F,\mathrm{ctrl}_F,\mathrm{link}_F):X\rightarrow Y$.
The concrete bigraph is written as $F=\langle F^{\mathsf{P}},F^{\mathsf{L}}\rangle$.
\end{definition}
Figure~\ref{bigraph::} shows a concrete bigraph $H$ that consists of the place graph shown in figure~\ref{place::graph}
and the link graph shown in figure~\ref{link::graph}.

The dynamics of bigraphs is defined in terms of rewrite rules that are known as {\em reaction rules}. 
These rules specify how bigraphs reconfigure themselves. In particular, a reaction rule specifies
a pattern that may be matched by a bigraph and how this should change any bigraph that matches it.
Stochastic bigraphs~\cite{krivine2008} and probabilistic bigraphs~\cite{benford2016} are bigraphs where 
reaction rules are associated with a {\em rate constant} and likelihood degree, respectively. However,
one should note that in the literature the terms ``stochastic'' and ``probabilistic'' tend to mean exactly 
the same thing. By replacing either the likelihood degrees or the rate constants with 
plausibility degrees, we obtain fuzzy bigraphs. From a syntactic point of view (i.e., how they look when 
we write them down on paper) there is no difference between stochastic, probabilistic, 
and fuzzy bigraphs. However, from a semantic point of view (i.e., what is the meaning of the numbers associated 
with rules and how each rule is chosen) there is a big difference between stochastic/probabilistic and fuzzy
bigraphs. However, I am not going to discuss fuzzy reaction rules here. Instead, I will discuss how to fuzzify
bigraphs themselves. 

\section{Fuzzifying Bigraphs}
In order to fuzzify bigraphs I will demonstrate how one can fuzzify its constituents, that is, how to fuzzify place 
and link graphs. This means that at least some parts of a fuzzy bigraph should be fuzzified.  In particular, the
mappings $\mathrm{ctrl}_{F}$, $\mathrm{prnt}_{F}$, and $\mathrm{link}_F$ will be replaced by the fuzzy mappings
$\widetilde{\mathrm{ctrl}}_{\widetilde{F}}$, $\widetilde{\mathrm{prnt}}_{\widetilde{F}}$, and 
$\widetilde{\mathrm{link}}_{\widetilde{F}}$. There are at least two methods to define fuzzy mappings. The first method 
is based on the remark that a function is actually a relation~\cite{chang1972}. Thus a fuzzy mapping $f$ from $X$ to $Y$ 
is a fuzzy set on $X\times Y$. A second approach is making use of the extension principle (e.g., see~\cite{buckley2002}) but 
in our case, fuzzy mappings that are fuzzy relations are quite adequate. 

\subsection{Fuzzy Bigraphs}
First, I need to describe fuzzy place graphs.
\begin{definition}
A fuzzy place graphs is a triple
\begin{displaymath}
\widetilde{F}=(V_{\widetilde{F}},\widetilde{\mathrm{ctrl}}_{\widetilde{F}},
                  \widetilde{\mathrm{prnt}}_{\widetilde{F}}):m\rightarrow n,
\end{displaymath}
where $\widetilde{\mathrm{ctrl}}_{\widetilde{F}}:V_{\widetilde{F}}\times\mathcal{K}\rightarrow L$ and 
$\widetilde{\mathrm{prnt}}_{\widetilde{F}}:(m\uplus V_{\widetilde{F}})\times (V_{\widetilde{F}}\uplus n)\rightarrow L$ 
are two $L$-fuzzy relations. Here $L$ is assumed to be frame.\footnote{A partially ordered set $P$ is a frame if and only if
\begin{enumerate}
\item every subset has a least upper bound;
\item every finite subset has a greatest lower bound; and 
\item the operator $\wedge$ distributes over $\vee$:
\begin{displaymath}
x\wedge\bigvee Y=\bigvee\left\{x\wedge y\mathrel{\bigm|}y\in Y \right\}.
\end{displaymath}
\end{enumerate}}
Usually, $L=[0,1]$ with the implied ordering. 
\end{definition}
The map $\widetilde{\mathrm{ctrl}}_{\widetilde{F}}$ specifies that a
given node has a number of ports with some plausibility degree. When we write $\widetilde{F}:m\rightarrow n$, we assume
that $\widetilde{F}$ has $m$ outer interfaces and $n$ inner interfaces but, in general, it is quite possible that some interfaces
are not really operational for any possible reason. Bigraphs are used to model existing systems and naturally
there are many systems that are far from being perfect. Thus the plausibility degree should be used to describe such 
{\em special} systems. In a similar way we define fuzzy link graphs:
\begin{definition}
A fuzzy link graphs is as a quadruple
\begin{displaymath}
\widetilde{F}=(V_{\widetilde{F}},E_{\widetilde{F}},\widetilde{\mathrm{ctrl}}_{\widetilde{F}},
               \widetilde{\mathrm{link}}_{\widetilde{F}}):X\rightarrow Y,
\end{displaymath}
where $\widetilde{\mathrm{ctrl}}_{\widetilde{F}}:V_{\widetilde{F}}\times\mathcal{K}\rightarrow L$ and 
$\widetilde{\mathrm{link}}_{\widetilde{F}}:(X\uplus P_{\widetilde{F}})\times (E_{\widetilde{F}}\uplus Y)\rightarrow L$ 
are two $L$-fuzzy relations. 
\end{definition}

Equipped with the definitions of fuzzy place and fuzzy link graphs, it trivial to give the
definition of fuzzy bigraphs.
\begin{definition}
A {\em concrete fuzzy bigraph} is a quintuple:
\begin{displaymath}
\widetilde{F}=(V_{\widetilde{F}},E_{\widetilde{F}},
               \widetilde{\mathrm{ctrl}}_{\widetilde{F}},
               \widetilde{\mathrm{prnt}}_{\widetilde{F}},
               \widetilde{\mathrm{link}}_{\widetilde{F}}):\langle k,X\rangle\rightarrow
                                                             \langle m,Y\rangle.
\end{displaymath}
\end{definition}
\subparagraph{Support of Fuzzy Bigraphs}
Suppose that $\widetilde{F}$ is fuzzy bigraph. Then, the support of $\widetilde{F}^{\mathsf{P}}$, denoted
$|\widetilde{F}|$, is the set $V_{\widetilde{F}}$. Also, the support of $\widetilde{F}^{\mathsf{L}}$
is the set $V_{\widetilde{F}}\uplus E_{\widetilde{F}}$. Further, assume that $\widetilde{F}$ and $\widetilde{G}$
are two fuzzy bigraphs that share the same sets of interfaces. Then, a support translation 
$\rho:|\widetilde{F}|\rightarrow|\widetilde{G}|$ consists of a pair of bijections
$\rho_V:V_{\widetilde{F}}\rightarrow V_{\widetilde{G}}$ and $\rho_E:E_{\widetilde{F}}\rightarrow E_{\widetilde{G}}$.
These bijections induce the $L$-fuzzy relations $\widetilde{\rho}_V:V_{\widetilde{F}}\times V_{\widetilde{G}}\rightarrow L$ 
and $\widetilde{\rho}_E:E_{\widetilde{F}}\times E_{\widetilde{G}}\rightarrow$ defined as follows:
\begin{displaymath}
\widetilde{\rho}_V(v,v')=\left\{ \begin{array}{ll}
                                 \top, & \text{if}\; \rho_V(v)=v'\\
                                 \bot, & \text{if}\; \rho_V(v)\neq v'\\
                                 \end{array}\right. \quad\text{and}\quad
\widetilde{\rho}_E(e,e')=\left\{ \begin{array}{ll}
                                 \top, & \text{if}\; \rho_E(e)=e'\\
                                 \bot, & \text{if}\; \rho_E(e)\neq e'\\
                                 \end{array}\right., 
\end{displaymath}
where $\top$ and $\bot$ are the top and bottom elements of $L$. These mappings should have the following properties:
\begin{enumerate}
\item $\widetilde{\mathrm{ctrl}}_{\widetilde{F}}\circ\widetilde{\rho}_V\le\widetilde{\mathrm{ctrl}}_{\widetilde{G}}$.
\item Map $\rho$ induces a bijection $\rho_P:P_{\widetilde{F}}\rightarrow P_{\widetilde{G}}$ such that 
$\rho_P((v,i)) \stackrel{\mathrm{def}}{=}(\rho_V(v),i)$. Clearly, this map induces the $L$-fuzzy relation
$\widetilde{\rho}_P:P_{\widetilde{F}}\times P_{\widetilde{G}}\rightarrow L$ defined as follows:
\begin{displaymath}
\widetilde{\rho}_P((v,i),(\rho_V(v),i))=\left\{\begin{array}{ll}
                                               \top, & \text{if}\; \rho_P((v,i))=(\rho_V(v),i),\\
                                               \bot, & \text{otherwise}.
                                               \end{array}\right.
\end{displaymath}
\item 
\begin{align*}
       \widetilde{\mathrm{prnt}}_{\widetilde{G}}\circ(\widetilde{\textsf{Id}}_m\uplus\widetilde{\rho}_V)&\ge
      (\widetilde{\mathsf{Id}}_n\uplus\widetilde{\rho}_V)\circ\widetilde{\mathrm{prnt}}_{\widetilde{F}}\\
       \widetilde{\mathrm{link}}_{\widetilde{G}}\circ(\widetilde{\textsf{Id}}_X\uplus\widetilde{\rho}_P)&\ge
      (\widetilde{\mathsf{Id}}_Y\uplus\widetilde{\rho}_E)\circ\widetilde{\mathrm{link}}_{\widetilde{F}}
\end{align*}
where $\widetilde{\mathsf{Id}}_m$ is a fuzzy $L$-relation produced from the identity function $\mathsf{Id}_m$.
\end{enumerate}

Although, I have promised not to discuss reaction rules, suffices to say that it is definitely possible to have 
fuzzy bigraphs with fuzzy reaction rules. In addition, being able only to modify bigraphs is not that useful. At
least, one should be able to compose bigraphs and create more complex structures. In fact, it is possible to
compose bigraphs  and so to define a category of whose arrows are bigraphs (see~\cite{milner2009} for the 
definition of bigraph composition). By extending the definition of bigraph 
composition, one can define the composition of fuzzy bigraph. It turns out that it is easier to define the composition 
of fuzzy plane graphs and the composition of fuzzy link graphs and based on these to define the composition of 
fuzzy bigraphs.
\subparagraph{Composition of Fuzzy Place Graphs}
Assume that $\widetilde{F}:k\rightarrow m$ and $\widetilde{G}:m\rightarrow n$ are two fuzzy place graphs such that
$|\widetilde{G}|\cap|\widetilde{F}|=\emptyset$. Then, the composite is the triple
\begin{displaymath}
\widetilde{G}\circ\widetilde{F}=(V,\widetilde{\mathrm{ctrl}},\widetilde{\mathrm{prnt}}):k\rightarrow n,
\end{displaymath} 
where $V=V_{\widetilde{F}}\uplus V_{\widetilde{G}}$, $\widetilde{\mathrm{ctrl}}=\widetilde{\mathrm{ctrl}}_{\widetilde{F}}\circ
\widetilde{ctrl}_{\widetilde{G}}$, and
\begin{displaymath}
\widetilde{prnt}(w,w')=\left\{ \begin{array}{ll}
                                 \widetilde{prnt}_{\widetilde{F}}(w,w'),& \text{if}\;w\in k\uplus V_{\widetilde{F}}\; \text{and}\;
                                 w'\in V_{\widetilde{F}},\\
                                 \widetilde{prnt}_{\widetilde{G}}(w,j), & \text{if}\;w\in k\uplus V_{\widetilde{F}}\; \text{and}\;
                                 j\in m,\\
                                 \widetilde{prnt}(w,w'), & \text{if}\;w\in \widetilde{prnt}_{\widetilde{G}}(w,w')
                                 \text{and}\; w\in V_{\widetilde{G}}.
                               \end{array}\right.                            
\end{displaymath}
The identity fuzzy place graph at $m$ is $\mathsf{id}_m=(\emptyset,\emptyset_{\mathcal{K}},\widetilde{\mathsf{Id}}_m):m\rightarrow m$.
\subparagraph{Composition of Fuzzy Link Graphs}
Suppose that $\widetilde{F}:X\rightarrow Y$ and $\widetilde{G}:Y\rightarrow Z$ are two link graphs such that 
$|\widetilde{F}|\cap|\widetilde{G}|=\emptyset$. Then, their composite is the link graph:
\begin{displaymath}
\widetilde{G}\circ\widetilde{F}=(V,E,\widetilde{\mathrm{ctrl}},\widetilde{\mathrm{link}}):X\rightarrow Z,
\end{displaymath}
where $V=V_{\widetilde{F}}\uplus V_{\widetilde{G}}$, $E=E_{\widetilde{F}}\uplus E_{\widetilde{G}}$,
\begin{displaymath}
\widetilde{\mathrm{ctrl}}(s,s')=\bigl(\widetilde{\mathrm{ctrl}}_{\widetilde{F}}\uplus\widetilde{\mathrm{ctrl}}_{\widetilde{G}}\bigr)(s,s')=
                                       \left\{\begin{array}{ll}
                                        \widetilde{\mathrm{ctrl}}_{\widetilde{F}}(s,s'), & \text{if}\; s,s'\in V_{\widetilde{F}},\\
                                        \widetilde{\mathrm{ctrl}}_{\widetilde{G}}(s,s'), & \text{if}\; s,s'\in V_{\widetilde{G}},
                                        \end{array}\right.
\end{displaymath}
and
\begin{displaymath}
\widetilde{\mathrm{link}}(q,q')=\left\{ \begin{array}{ll}
                                        \widetilde{\mathrm{link}}_{\widetilde{F}}(q,q'), & \text{if}\; q\in X\uplus P_{\widetilde{F}}\;
                                        and\; q'\in E_{\widetilde{F}},\\
                                        \widetilde{\mathrm{link}}_{\widetilde{G}}(q,y), & \text{if}\; q\in X\uplus P_{\widetilde{F}}\;
                                        and\; y\in Y,\\
                                        \widetilde{\mathrm{link}}_{\widetilde{G}}(q,q'), & \text{if}\; q\in P_{\widetilde{G}}.
                                        \end{array}\right. 
\end{displaymath}
The identity fuzzy link graph at $X$ is $\mathsf{id}_X=(\emptyset,\emptyset,\emptyset_{\mathcal{K}},\widetilde{\mathsf{Id}}_X):X\rightarrow X$.
\subparagraph{Composition of Fuzzy Bigraphs}
If $\widetilde{F}:I\rightarrow J$ and $\widetilde{G}:J\rightarrow K$ are two fuzzy bigraphs, such that 
$|\widetilde{F}|\cap|\widetilde{G}=\emptyset$, their composite is the pair
\begin{displaymath}
\widetilde{G}\circ\widetilde{F}=(\widetilde{G}^{\mathsf{P}}\circ\widetilde{F}^{\mathsf{P}},
                                 \widetilde{G}^{\mathsf{L}}\circ\widetilde{F}^{\mathsf{L}}):I\rightarrow K
\end{displaymath}
and the identity fuzzy bigraph at $I=\langle m,X\rangle$ is $\langle\mathsf{id}_m,\mathsf{id}_X\rangle$.
\begin{theorem}
Given three fuzzy bigraphs $A:I\rightarrow J$, $B:J\rightarrow K$, and $C:K\rightarrow M$, then
\begin{displaymath}
C\circ(B\circ A)=(C\circ B)\circ A.
\end{displaymath}
\end{theorem}
The proof is based on the solution of exercise 2.1 in Milner's book~\cite{milner2009} and the fact that
composition of fuzzy relations is associative.
\subparagraph{Tensor Product of Fuzzy Bigraphs}
Given two disjoint fuzzy place graphs $\widetilde{F}:k\rightarrow l$ and $\widetilde{G}:m\rightarrow n$, 
their tensor product is the fuzzy place graph $\widetilde{F}\otimes\widetilde{G}:k+m\rightarrow l+n$ defined as follows:
\begin{displaymath}
\widetilde{F}\otimes\widetilde{G}\stackrel{\mathrm{def}}{=}(V_{\widetilde{F}}\uplus V_{\widetilde{G}},
                                           \widetilde{\mathrm{ctrl}}_{\widetilde{F}}\uplus\widetilde{\mathrm{ctrl}}_{\widetilde{G}}, 
                                           \widetilde{\mathrm{prnt}}_{\widetilde{F}}\uplus\widetilde{\mathrm{prnt}}'_{\widetilde{G}}),              
\end{displaymath}
where $\widetilde{\mathrm{prnt}}'_{\widetilde{G}}(k+i,l+i)\ge\widetilde{\mathrm{prnt}}_{\widetilde{G}}(i,j)$. The unit of $\otimes$ is
$0$.

Given two fuzzy link graphs $\widetilde{F}:X\rightarrow Y$ and $\widetilde{G}:W\rightarrow Z$, then their tensor product is the
fuzzy link graph $\widetilde{F}\otimes\widetilde{G}:X\uplus W\rightarrow Y\uplus Z$ defined as follows:
\begin{displaymath}
\widetilde{F}\otimes\widetilde{G}\stackrel{\mathrm{def}}{=}(V_{\widetilde{F}}\uplus V_{\widetilde{G}},
                                                            E_{\widetilde{F}}\uplus E_{\widetilde{G}},
                                           \widetilde{\mathrm{ctrl}}_{\widetilde{F}}\uplus\widetilde{\mathrm{ctrl}}_{\widetilde{G}}, 
                                           \widetilde{\mathrm{link}}_{\widetilde{F}}\uplus\widetilde{\mathrm{link}}_{\widetilde{G}}).              
\end{displaymath}
The unit of the tensor product of fuzzy link graphs is $\emptyset$. It is now obvious what is the tensor product of two fuzzy
bigraphs. The unit of the tensor product for fuzzy bigraphs is $\epsilon=\langle 0,\emptyset\rangle$.
\subsection{Type~2 Fuzzy Bigraphs}
It is quite possible to have ``fuzzier'' bigraphs by fuzzifying the sets $V_{\widetilde{F}}$ and $E_{\widetilde{F}}$.
Thus the set $V_{\widetilde{F}}$ will be replaced by the fuzzy set 
$\widetilde{V}_{\widetilde{F}}:\mathcal{V}\rightarrow L$ and the fuzzy set $\widetilde{E}_{\widetilde{F}}$ will be
replaced by the fuzzy set  $\widetilde{E}_{\widetilde{F}}:\mathcal{E}\rightarrow L$. The meaning 
of these fuzzy sets is that nodes are part of a forest to some degree and edges ``exist'' up to some degree because
connections are broken, etc. 
\begin{definition}
A {\em type~2 fuzzy place graph} is 
a triple:
\begin{displaymath}
\widetilde{F}=(\widetilde{V}_{\widetilde{F}},\widetilde{\mathrm{ctrl}}_{\widetilde{F}},
\widetilde{\mathrm{prnt}}_{\widetilde{F}}):m\overset{\beta}{\rightarrow}n
\end{displaymath}
where  $\widetilde{\mathrm{ctrl}}_{\widetilde{F}}:\mathcal{V}\times\mathcal{K}\rightarrow L$ and
$\widetilde{\mathrm{prnt}}_{\widetilde{F}}:\mathcal{V}\times\mathcal{V}\rightarrow L$ are two $L$-fuzzy relations,
and $\beta\in L$ specifies the degree to which the type~2 fuzzy place graph has $m$ (functional) inner interfaces 
and $n$ (functional) outer interfaces. 
\end{definition}
Similarly, one can define type~2 fuzzy link graph as follows:
\begin{definition}
A {\em type~2 fuzzy link graph} is a quadruple
\begin{displaymath}
\widetilde{F}=(\widetilde{V}_{\widetilde{F}},\widetilde{E}_{\widetilde{F}},
\widetilde{\mathrm{ctrl}}_{\widetilde{F}},\widetilde{\mathrm{link}}_{\widetilde{F}}):X\overset{\delta}{\rightarrow} Y,
\end{displaymath}
where $\widetilde{\mathrm{ctrl}}_{\widetilde{F}}:\mathcal{V}\times\mathcal{K}\rightarrow L$, 
$\widetilde{\mathrm{link}}_{\widetilde{F}}:(X\uplus P_{\widetilde{F}})\times\mathcal{E}\rightarrow L$ are 
two $L$-fuzzy relations, $\delta\in L$ is the degree to which the type~2 fuzzy link graph has $\card X$ (functional) inner 
interfaces and $\card Y$ (functional) outer interfaces. Here the set $P_{\widetilde{F}}$ is defined as follows:
\begin{displaymath}
P_{\widetilde{F}} \stackrel{\mathrm{def}}{=} \Bigl\{(v,i)\mathrel{\Bigm|}
\bigvee_{v}\bigvee_{k}\widetilde{\mathrm{crtl}}_{\widetilde{F}}(v,k)\;\text{and}\; i\in\mathrm{ar}(k) \Bigr\}.
\end{displaymath}
Note that here we examine all pairs $(v,k)$ and chose the one that can be used to compute the infimum and 
then use this $k$ to compute $\mathrm{ar}(k)$. 
\end{definition}

Equipped with these definition, it is straightforward to formulate the definition of a 
{\em concrete fuzzy bigraph} as a quintuple:
\begin{displaymath}
\widetilde{F}=(\widetilde{V}_{\widetilde{F}},\widetilde{E}_{\widetilde{F}},
               \widetilde{\mathrm{ctrl}}_{\widetilde{F}},
               \widetilde{\mathrm{prnt}}_{\widetilde{F}},
               \widetilde{\mathrm{link}}_{\widetilde{F})}:\langle k,X\rangle\overset{\gamma}{\rightarrow}
                                                             \langle m,Y\rangle,
\end{displaymath}
where $\gamma=\beta\wedge\delta$.
\subparagraph{Support of Type~2 Fuzzy Bigraphs}
For a type 2 fuzzy place graph $\widetilde{G}$ its support $|\widetilde{G}|$ is $\widetilde{V}_{\widetilde{G}}$ and for a
type 2 fuzzy link graph or a type 2 fuzzy bigraph $\widetilde{G}$ its support $|\widetilde{G}|$ is $\widetilde{V}_{\widetilde{G}}\uplus
\widetilde{E}_{\widetilde{G}}$, where
\begin{displaymath}
\bigl(\widetilde{V}_{\widetilde{G}}\uplus\widetilde{E}_{\widetilde{G}}\bigr)(v,e)=
\widetilde{V}_{\widetilde{G}}(v)\vee\widetilde{E}_{\widetilde{G}}(e).
\end{displaymath}
Note that $\widetilde{V}_{\widetilde{G}}(e)=\bot$ and $\widetilde{E}_{\widetilde{G}}(v)=\top$ and so
I ``simulate'' the functionality of $V_{\widetilde{F}}\uplus E_{\widetilde{G}}$.

A support translation $\widetilde{\rho}:|\widetilde{F}|\rightarrow|\widetilde{G}|$ from 
$\widetilde{F}$ to $\widetilde{G}$ consists of pair of fuzzy relations 
$\widetilde{\rho}_V:\mathcal{V}\times\mathcal{V}\rightarrow L$ and 
$\widetilde{\rho}_E:\mathcal{E}\times\mathcal{E}\rightarrow L$ such that
\begin{displaymath}
\widetilde{\rho}_V(v,v')=\widetilde{V}_{\widetilde{G}}(v')\quad\text{and}\quad
\widetilde{\rho}_E(e,e')=\widetilde{E}_{\widetilde{G}}(e').
\end{displaymath}
Moreover, $\widetilde{\rho}$ {\em preserves} controls, that is, 
$\widetilde{\mathrm{crtl}}_{\widetilde{G}}\circ\widetilde{\rho}_{V}\le\widetilde{\mathrm{ctrl}}_{\widetilde{F}}$.
Also, $\widetilde{\rho}_V$ induces another fuzzy relation 
$\widetilde{\rho}_P:P_{\widetilde{F}}\times P_{\widetilde{G}}\rightarrow L$ such that
\begin{displaymath}
\widetilde{\rho}_V(v,v')\le \widetilde{\rho}_P\bigl((v,i),(v',i')\bigr). 
\end{displaymath}
In addition, the following inequalities should hold:
\begin{align*}
\widetilde{\mathrm{prnt}}_{\widetilde{G}}\circ(\mathrm{Id}_{m}\uplus\widetilde{\rho}_V) &\le
(\mathrm{Id}_n\uplus\widetilde{\rho}_V)\circ\widetilde{\mathrm{prnt}}_{\widetilde{F}}\\
\widetilde{\mathrm{link}}_{\widetilde{G}}\circ(\mathrm{Id}_{X}\uplus\widetilde{\rho}_V) &\le
(\mathrm{Id}_n\uplus\widetilde{\rho}_V)\circ\widetilde{\mathrm{prnt}}_{\widetilde{F}}
\end{align*}
Here $\mathrm{Id}_m:m\times m\rightarrow L$ is a map such that $\mathrm{Id}_m(m,m)=\top$ and
$\mathrm{Id}_m(n,m)=\bot$, respectively. Similar definitions hold for $\mathrm{Id}_n$, $\mathrm{Id}_X$, 
and $\mathrm{Id}_Y$.
Given $\widetilde{F}$ and $\rho$ we can determine $\widetilde{G}$. When this happens, then we
say that they are {\em support equivalent} and write  $\widetilde{F}\bumpeq\widetilde{G}$.
\subparagraph{Composition of Type~2 Fuzzy Bigraphs}
Composition of type 2 fuzzy bigraphs is defined pairwise. First, if $\widetilde{F}:k\overset{\mu}{\rightarrow} m$ and 
$\widetilde{G}:m\overset{\nu}{\rightarrow}n$ are two type~2 fuzzy place graphs, then the following triple is the
composite type~2 fuzzy place graph: 
\begin{displaymath}
\widetilde{G}\circ\widetilde{F}=(\widetilde{V},\widetilde{\mathrm{ctrl}},
\widetilde{\mathrm{prnt}}):k\overset{\kappa}{\rightarrow} m,
\end{displaymath}
where  $\kappa=\mu\wedge\nu$, $\widetilde{V}(v)=\bigl(\widetilde{V}_{\widetilde{F}}\uplus\widetilde{V}_{\widetilde{G}}\bigr)(v)$,
$\widetilde{\mathrm{ctrl}}=\widetilde{\mathrm{ctrl}}_{\widetilde{F}}\uplus\widetilde{\mathrm{ctrl}}_{\widetilde{G}}$, and 
\begin{displaymath}
\widetilde{\mathrm{prnt}}(w,w') =\left\{ \begin{array}{ll}
               \widetilde{\mathrm{prnt}}_{\widetilde{F}}(w,w'), & \text{if}\; (k\uplus\widetilde{V}_{\widetilde{F}})(w)\ge\kappa
                \;\text{and}\;\widetilde{prnt}_{\widetilde{F}}(w,w')\ge\kappa,\\
               \widetilde{\mathrm{prnt}}_{\widetilde{G}}(w',j), & \text{if}\; (k\uplus\widetilde{V}_{\widetilde{G}})(w)\ge\kappa
                 \;\text{and}\;\widetilde{prnt}_{\widetilde{G}}(w,j)\ge\kappa,\\
               \widetilde{\mathrm{prnt}}_{\widetilde{G}}(w,w'), & \text{if}\;\widetilde{V}_{\widetilde{G}}(w)\ge\kappa.
                                         \end{array}\right.
\end{displaymath}
Also, if $\widetilde{F}:X\overset{\mu}{\rightarrow} Y$ and $\widetilde{G}:Y\overset{\nu}{\rightarrow}Z$ are two type~2 fuzzy 
link graphs, then the following quadruple is the composite type~2 fuzzy link graph: 
\begin{displaymath}
\widetilde{G}\circ\widetilde{F}=(\widetilde{V},\widetilde{E},\widetilde{\mathrm{ctrl}},
\widetilde{\mathrm{link}}):X\overset{\kappa}{\rightarrow} Y,
\end{displaymath}
where  $\widetilde{E}(e)=\bigl(\widetilde{E}_{\widetilde{F}}\uplus\widetilde{E}_{\widetilde{G}}\bigr)(e)$ and
\begin{displaymath}
\widetilde{\mathrm{link}}(q,q') =\left\{ \begin{array}{ll}
               \widetilde{\mathrm{link}}_{\widetilde{F}}(q,q'), & \text{if}\; q\in (X\uplus P_{\widetilde{F}})
                \;\text{and}\;\widetilde{link}_{\widetilde{F}}(q,q')\ge\kappa,\\
               \widetilde{\mathrm{link}}_{\widetilde{G}}(q',y), & \text{if}\; q\in (X\uplus P_{\widetilde{G}})(q)
                 \;\text{and}\;\widetilde{link}_{\widetilde{G}}(q,j)\ge\kappa,\\
               \widetilde{\mathrm{link}}_{\widetilde{G}}(q,q'), & \text{if}\; q\in P_{\widetilde{G}}(q).
                                         \end{array}\right.
\end{displaymath}
From here it is a straightforward exercise to define the composition of type~2 fuzzy bigraphs. Also, one can prove that
composition is an associative operation. The identities are the identities of fuzzy bigraphs and their plausibility
degree is equal to $\top$.
\section{Fuzzy Bigraphical Categories}
Usually, when we define a category, first we define its objects and then the arrows between objects. However, in the
case if fuzzy bigraphical categories I have already introduced arrows in the previous section. Fuzzy place graphs, 
fuzzy link graphs, and fuzzy bigraphs are arrows that can be composed also but not all compositions are possible.
If fuzzy place graphs, fuzzy link graphs, and fuzzy bigraphs are the arrows of different categories, what are the
objects of these categories? The answer is very simple: The objects of these categories are natural numbers, finite sets 
of symbols, and pairs of a natural number and a finite set of symbols, respectively. For type~2 fuzzy bigraphs, we need
a new kind of category where each arrow is associated with a plausibility degree. The following definition introduces such
a new kind of category theory (see~\cite{syropoulos2014b} for more details).
\begin{definition}
A fuzzy category $\widetilde{\mathcal{C}}$ is an ordinary category $\mathcal{C}$ but in addition: 
\begin{enumerate}
\item There is an operation $\mathop{\mathrm{p}}$ that assigns to each arrow a plausibility degree
$\rho=\mathop{\mathrm{p}}(f)\in [0,1]$. Thus an arrow that starts from $A$ and ends at $B$ with 
plausibility degree $\rho$ is written as:
\begin{displaymath}
A\overset{f}{\underset{\rho}{\longrightarrow}}B\quad\mbox{or}\quad
f:A\overset{\rho}{\rightarrow}B;
\end{displaymath}
\item For the {\em composite} $g\circ f$ it holds that
 $\mathop{\mathrm{p}}(g\circ f)=\mathop{\mathrm{p}}(f)\wedge  \mathop{\mathrm{p}}(g)$. The
associative law holds since $\wedge$ is an associative operation.
\item an {\em assignment} to each $\widetilde{\mathcal{C}}$-object $B$ of a $\widetilde{\mathcal{C}}$-arrow
$\mathbf{1}_{B}:B\overset{1}{\longrightarrow}B$, called the {\em identity arrow on} $B$, such
that the following {\em identity law} holds true:
\begin{displaymath}
\mathbf{1}_B\circ f=f\quad\mbox{and}\quad g\circ\mathbf{1}_B=g
\end{displaymath} 
for any $\widetilde{\mathcal{C}}$-arrows $f:A\overset{\rho_f}{\longrightarrow}B$ and
$g:B\overset{\rho_g}{\longrightarrow}A$.
\end{enumerate}
\end{definition}
Thus the following type-2 fuzzy bigraph
\begin{displaymath}
\widetilde{F}=(\widetilde{V}_{\widetilde{F}},\widetilde{E}_{\widetilde{F}},
               \widetilde{\mathrm{ctrl}}_{\widetilde{F}},
               \widetilde{\mathrm{prnt}}_{\widetilde{F}},
               \widetilde{\mathrm{link}}_{\widetilde{F})}:\langle k,X\rangle\overset{\gamma}{\rightarrow}
                                                             \langle m,Y\rangle
\end{displaymath}
is a fuzzy arrow from $\langle k,X\rangle$ to $\langle m,Y\rangle$ with plausibility degree $\gamma$:
\begin{displaymath}
\langle k,X\rangle\overset{\widetilde{F}}{\underset{\gamma}{\longrightarrow}}\langle m,Y\rangle.
\end{displaymath}

\section{Conclusions}
I have introduced fuzzy bigraphs and type~2 fuzzy bigraphs. I have described how one can compose fuzzy bigraphs and
type~2 fuzzy bigraphs. Also, I described how to define a category whose arrows are fuzzy bigraphs and I introduced
a fuzzy version of a category in order to be able to define a fuzzy category whose objects are type~2 fuzzy bigraphs.
Naturally, there are many things to be done in order to have a fully fledged theory of fuzzy ubiquitous computing 
but this is a task that requires only time\ldots

\end{document}